\documentclass[sigconf]{acmart}
\usepackage{url}
\usepackage{xargs}                      
\usepackage[colorinlistoftodos,prependcaption,textsize=tiny]{todonotes}
\usepackage{graphicx}
\usepackage{longtable,lscape}
\usepackage{forest}
\usepackage{tikz-qtree}
\usepackage{array}
\usepackage{cleveref}
\usepackage[inline]{enumitem}   
\usepackage{booktabs}
\usepackage{multirow}
\usepackage{tabularx} 
\usepackage{collcell}
\usepackage{rotating}
\usepackage{geometry}
\usepackage{adjustbox}
\usepackage{pgfplots}
\usepackage{listings}
\usepackage{amsmath}
\usepackage{subcaption}
\usepackage{numprint}
\usepackage{makecell}
\usepackage{flushend}
\usepackage{threeparttable}
\usepackage[nomargin,inline,marginclue,draft]{fixme}
\hypersetup{draft} 

\newcommandx{\unsure}[2][1=]{\todo[linecolor=red,backgroundcolor=red!25,bordercolor=red,#1]{#2}}
\newcommandx{\change}[2][1=]{\todo[linecolor=blue,backgroundcolor=blue!25,bordercolor=blue,#1]{#2}}
\newcommandx{\info}[2][1=]{\todo[linecolor=OliveGreen,backgroundcolor=OliveGreen!25,bordercolor=OliveGreen,#1]{#2}}
\newcommandx{\improvement}[2][1=]{\todo[linecolor=Plum,backgroundcolor=Plum!25,bordercolor=Plum,#1]{#2}}
\newcommandx{\thiswillnotshow}[2][1=]{\todo[disable,#1]{#2}}

\newcommand*\rot{\rotatebox{90}}
\definecolor{dkgreen}{rgb}{0,0.6,0}
\definecolor{gray}{rgb}{0.5,0.5,0.5}
\definecolor{dkgreen}{rgb}{0,0.6,0}
\definecolor{darkgreen}{RGB}{0,102,0}
\definecolor{gray}{rgb}{0.5,0.5,0.5}
\definecolor{mauve}{rgb}{0.58,0,0.82}
\crefname{lstlisting}{listing}{listings}
\Crefname{lstlisting}{Listing}{Listings}

\setcopyright{acmcopyright}

\copyrightyear{2019}
\acmYear{2019}
\acmConference[SPRO'19]{3rd Software Protection Workshop}{November 15, 2019}{London, United Kingdom}
\acmBooktitle{3rd Software Protection Workshop (SPRO'19), November 15, 2019, London, United Kingdom}
\acmPrice{15.00}
\acmDOI{10.1145/3338503.3357723}
\acmISBN{978-1-4503-6835-3/19/11}

\begin{document}

\fancyhead{}
\title{VirtSC: Combining Virtualization Obfuscation with Self-Checksumming}
%
%
%
%
%

\author{Mohsen Ahmadvand, Daniel Below, Sebastian Banescu, and Alexander Pretschner}
\orcid{1234-5678-9012}
\affiliation{%
	\institution{Technical University of Munich}
}
\email{firstname.lastname@cs.tum.edu}

\renewcommand{\shortauthors}{Mohsen Ahmadvand et al.}

\begin{abstract}
Self-checksumming (SC) is a tamper-proofing technique that ensures certain program segments (code) in memory hash to known values at runtime.
SC has few restrictions on application and hence can protect a vast majority of programs.
The code verification in SC requires computation of the expected hashes after compilation, as the machine-code is not known before. 
This means the expected hash values need to be adjusted in the binary executable, 
hence combining SC with other protections is limited due to this adjustment step.
However, obfuscation protections are often necessary, as SC protections can be otherwise easily detected and disabled via pattern matching. 
In this paper, we present a layered protection using virtualization obfuscation, yielding
an architecture-agnostic SC protection that requires no post-compilation adjustment.
We evaluate the performance of our scheme using a dataset of 25 real-world programs (MiBench and 3 CLI games). 
Our results show that the SC scheme induces an average overhead of 43\% for a complete protection (100\% coverage).
The overhead is tolerable for less CPU-intensive programs (e.g. games) and when only parts of programs (e.g. license checking) are protected.
However, large overheads stemming from the virtualization obfuscation were encountered. 

\end{abstract}


\maketitle              

\section{Introduction}\label{sec:introduction}
Man-at-the-end (\emph{MATE}) attacks are a threat to software execution integrity and intellectual property of the entity that developed the software. 
Unprotected software that is distributed to end-users makes it possible for malicious end-users to tamper with it (also called \textit{code manipulation}) both statically and dynamically, because they can control the execution environment. 
Dynamic tampering can be achieved by changing the memory of a running program,
for example by attaching a debugger to the process. Tampering with software in a static context is achieved by altering specific bytes of the executable. 
If performed correctly, this can cause the program to divert from its intended behavior and potentially offer premium features at no cost. 
Such a behavioral change could for example be the program skipping license key validation, 
making way for illegal distribution. 
The 2018 Global Software Survey\footnote{https://gss.bsa.org/} by the BSA, indicates that 37\% of software installed on personal computers is unlicensed.  
In BRIC countries the share of pirated software is estimated to be twice as large at around 60\%~\cite{bsagssreport2017,shareunlicensed2009to2017}. 
This constitutes an economic loss to software developers and it should not only be dealt with by legal means. 
It is therefore of interest to software developers to use software protection techniques to hamper code tampering and reduce illegal distribution.

\textbf{Problem.} 
Self-checksumming (SC) is a software protection technique that allows programs to detect changes in their binary representation and memory using so called \emph{guards}~\cite{chang2001protecting}. 
Upon detection, SC may call a response mechanism, for example aborting the execution or self-repair. 
However, since these checks have to be executed at runtime, 
the expected checksum values need to be pre-computed and, after compilation, inserted into the executable. 
Not only does this approach require knowledge of the underlying system's architecture, 
it also mandates a post-patching step to put these expected checksums (hashes) into predefined places.
This is an extremely tedious and error-prone process~\cite{banescu2017detecting}. 
This process also limits the use of other obfuscation techniques, as one needs to maintain a set of placeholders (i.e.~contiguous sequences of 4 or 8 bytes in the code segment) for the pre-computed checksums at known offsets in the executable.
Applying obfuscation would likely change the offsets and contiguous layout of the placeholders.

To eliminate this post-patching step, we leverage the compiler framework LLVM~\cite{lattner2004llvm} to implement self-checksumming atop virtualized instructions. For this, we also implement virtualization obfuscation. LLVM already implements backends for different system architectures, which removes the required architecture knowledge and post-patching.
Applying the guards at a higher abstraction level removes the post-patching process at binary level entirely. 
This is achieved by first applying virtualization obfuscation~\cite{ghosh2010secure} 
and then adding the guards in the custom interpreter bytecode. 

\textbf{Contributions.} This work makes the following contributions:
\begin{itemize}
	\item A novel design for combining self-checksumming and virtualization obfuscation (Section~\ref{sec:design}). 
	\item A performance evaluation of the implementation, using a dataset of 24 real-world programs (Section~\ref{chap:evaluation}).
	\item An attack-tree based security analysis of the design and implementation (Section~\ref{chap:evaluation}).
\end{itemize}
The rest of this paper is organized as follows.
Section~\ref{sec:background-and-related-work} presents the necessary background knowledge for this paper and related work.
Section~\ref{sec:discussion} discusses performance and security tradeoffs.
Section~\ref{sec:limitations} presents the limitations of this approach and implementation.
Finally, the conclusions are presented in Section~\ref{sec:conclusions}.

\section{Background and Related Work}\label{sec:background-and-related-work}

In this section we first present the two essential parts of our approach, i.e.~\emph{virtualization obfuscation} and \emph{self-checksumming}.
Afterwards, we present related work about combining the two techniques.

\subsection{Virtualization Obfuscation (VO)}\label{subsec:2-virtualization-obfuscation}

VO's primary goal is to transform a program's control flow to a semantically equivalent, yet less comprehensible version.
Given a program, this technique lifts all instructions to a new, random \emph{Instruction Set Architecture} (ISA). 
Thereafter, an interpreter specific to the newly generated ISA is created.
Lifted code along with the interpreter are what the end user receives.

The interpreter's job is to fetch, decode and dispatch execution to the original instructions' \textit{handler}. 
Each handler emulates the original instructions' behavior. 
Furthermore, all program memory allocations are done via a \emph{virtual memory} (VM).
A \emph{virtual program counter} (VPC) keeps track of the last executed instruction at runtime.
This obfuscation technique can be applied at different representations of programs such as source code~\cite{collberg2015tigress} or binary level~\cite{linn2003obfuscation}.
The level at which the technique is applied plays an important role in its composability with other protections.
To the best of our knowledge, there is no publicly available implementation of VO as a compiler pass particularly in LLVM. 

VO intuitively reduces the comprehensibility of protected programs for attackers.
VO can resists automated attacks imposing additional cost~\cite{anckaert2006proteus,cazalas2014probing,banescu2016vot4cs} at the perpetrator's end. 
For instance, the utilization of opaque predicates~\cite{collberg1998manufacturing} or range dividers~\cite{banescu2016code} can significantly hamper attacks based on symbolic execution~\cite{salwan2018symbolic}.

\subsection{Self-Checksumming (SC)}\label{subsec:2-self-checksumming}

Self-checksumming is a software tamper-proofing technique~\cite{chang2001protecting}. 
The idea is to equip a software with a set of interconnected guards.
Each of these guards carries out hash calculations over the code segment (in the process memory) during runtime to detect potential code manipulations.
Guards need to be pre-seeded with the expected hash values of the code that they are protecting. 
Upon detection of code tampering (i.e.~hash mismatch), a response mechanism is triggered~\cite{nagra2009surreptitious}. 

Since all calculations are done at runtime over the binary (machine code), the expected hashes can only be known after compilation of programs.
That is, if an SC protection is to be developed as a compiler pass, 
the expected hashes need to be adjusted once the binary representation of the program is finalized. 
For this purpose usually placeholders along with post-patching mechanisms are utilized~\cite{ahmadvand2018practical}.
Alternatively, a backend pass (similar to the one described in~\cite{junod2015obfuscator}) can be used to adjust placeholders.

Both of the mentioned adjustment approaches have a tight dependency on the underlying architecture. 
That is, the adjustment shall be tailored for each and every architecture for which the binary is compiled.
The architectural dependency imposes extra cost in terms of development and maintenance of protections.

Furthermore, SC guards are susceptible to pattern matching attacks~\cite{banescu2017detecting,ahmadvand2018practical,ahmadvand2019taxonomy}. 
Therefore, without proper utilization of obfuscations, SC is rather easily identifiable, and perhaps defeatable. 
However, obfuscation inherently alters the syntactical representation of programs. 
This comes with two drawbacks in the case of SC protection: 
\textbf{i)} the placeholders need to be preserved (not obfuscated) otherwise some adjustments may fail; and
\textbf{(ii)} obfuscating programs (after SC is applied) could potentially break SC guards (as expected hashes may no longer hold true). 
These setbacks may result in having no obfuscation on overlapping guards and expected hash values (placeholders before adjustments), which in turn negatively impacts the resilience of SC. 
As a direct consequence, the composability of SC with other protections is heavily limited and thus the overall security. 
In this paper we propose a technique to overcome the identified drawbacks and to fix the composability problem of SC. 

\subsection{Combining Virtualization and Integrity Protection}
Since our proposal is based on combining VO and SC, 
we reviewed existing literature on protection schemes that use VO to add resilience to their schemes.

Ghosh et al. \cite{ghosh2010secure} proposed a protection technique that combines process virtualization (comparable to VO), encryption, and self-checksumming. 
In their technique, program instructions together with their protection guards are encrypted and subsequently shipped into the binary. 
The decryption key is also placed in the binary but protected using white-box cryptography~\cite{chow2002white}. 
The process virtualization is rather used as a proxy to decrypt protected instructions. 
Since the SC protection is directly applied on the instructions, 
utilizing further protections (after the application of SC) could break the SC protection.
In another similar work Ghosh et al. \cite{ghosh2013software} used SC to protect dynamically generated (cached) instructions of the virtualized programs.

Gan et al.~\cite{gan2015using} proposed a technique that uses protected virtual machines acting as the root of trust. 
Several hardening techniques including white-box cryptography, obfuscations, etc. are also utilized to protect the virtual machine. 
The authors indicate such a virtual machine can be relied on to carry out integrity checks. 


In contrast to the related work, our proposal rather aims at improving composability of protections and
removing architectural dependencies and thus reducing costs.

\section{Design}
\label{sec:design}
In this section we present the architecture of our approach, the design decisions and the reasoning behind them. We end this section describing the optimizations that were added to the design.
\subsection{High-level Architecture}\label{sec:scvirt-workflow}
\begin{figure}[!tbp]
	\centering
	\includegraphics[clip, trim=0.25cm .375cm 0.3cm 0.2cm,width=\linewidth]{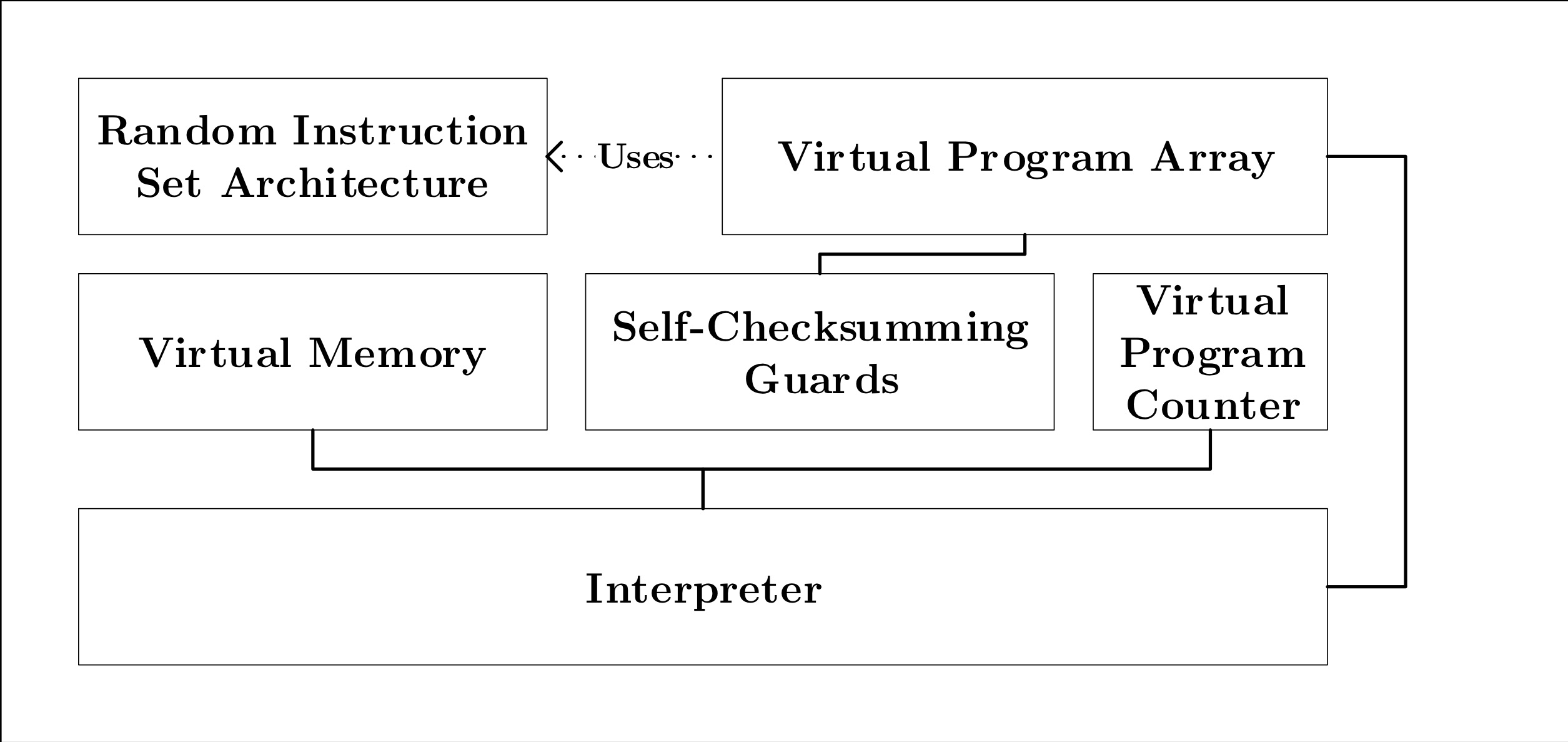}
	\caption{Solution's high level architecture}\label{fig:high-level-architecture}
\end{figure}

The architecture of a protected program using our technique is presented in \Cref{fig:high-level-architecture}.
Every program utilizes its own \emph{Random (Virtual) Instruction Set Architecture} (RISA). 
The RISA (depicted in the top-left corner of Figure~\ref{fig:high-level-architecture}), is a volatile data structure capturing a set of random instruction mnemonics and their actual semantics. 
The RISA only exists at the time of transformation and it is discarded afterwards.
The RISA is created sequentially while visiting program instructions. 

Our technique can be applied on either the entire application or a subset of its functions. 
Users can annotate the desired functions with a \emph{sensitive} tag. 
By iterating over instructions (of annotated functions) a \texttt{Virtual Program Array} (VPA) containing the translation of the original instruction to the new RISA is created (top-right of Figure~\ref{fig:high-level-architecture}).  
That is, if an instruction does not have an equivalent mnemonic in RISA, a corresponding mnemonic and a reference to its semantics are inserted into RISA. 
Next, all memory accesses (constants, variables and registers) are altered to use a sequential \texttt{Virtual Memory} (VM), which is depicted in the middle-left side of Figure~\ref{fig:high-level-architecture}. 
In addition to mnemonics, which are captured as random opcodes, the VPA keeps a reference to the indexes at which corresponding data of instructions are persisted in the VM.  

At this stage, self-checksumming guards according to the specified protection requirements are injected into the VPA (center of Figure~\ref{fig:high-level-architecture}). 
Guards, similar to the original program instructions, use the VM to carry out their calculations and subsequently verifications. 
Simply put, the computed hash as well as the expected hash values are stored in and retrieved from the VM. 
It is important to note that SC guards protect the VPA; 
i.e. they compute hashes over the VPA (not the original program instructions). 

For the lifted instructions to be executable an \texttt{Interpreter} is needed (bottom of Figure~\ref{fig:high-level-architecture}). 
The Interpreter fetches instructions from the VPA. It decodes the RISA's semantic of mnemonics to execute the fetched instructions. 
Within the interpreter a \texttt{Virtual Program Counter} (VPC) keeps track of the last executed instruction (see middle-right part of Figure~\ref{fig:high-level-architecture}).

\subsection{Detailed Design}
This section presents several architectural components from Section~\ref{sec:scvirt-workflow}, and their inner workings in more detail.

\subsubsection{Random Instruction Set Architecture (RISA)}\label{subsec:design-RISA}
The RISA is a program/function specific contract comprised of \emph{Opcode}, \emph{Operand(s)}, and \emph{Semantics}. 
The opcode, in essence, captures mnemonics of the RISA. Opcodes are randomly generated 16-bit integers. 
Operands refer to indexes in the VM where either the input or output of the computation should be ``retrieved from'' or ``stored to''. 

\subsubsection{Virtual Memory (VM)}\label{subsubsec:data-array}
The VM holds all operands needed by the translated instructions. 
Values are stored contiguously, and accessed by index. 
As types are known beforehand, we can deterministically compute the amount of elements we need to read or write upon load and store operations, respectively.
To do so, we simply set a pointer to the index corresponding to the ``beginning of'' and ``operand of'' interest followed with a cast to the known type. 
After dereferencing this pointer, the value can be used by the interpreter.
Persisting changes to VM values includes splitting the data into chunks of 8-bit integer values.

\subsubsection{Virtual Program Array (VPA)}\label{subsubsec:code-array}
The VPA is a static array containing the lifted instructions. 
As we are using an interpreter per-function (see section~\ref{subsubsec:interpreter}), 
each function has its own VPA and its checksum can be verified by any other function in the same module. 
In the case of a control-flow changing instruction the next VPC value is provided in the bytecode, which the interpreter will assign to the VPC.
Since we are using 16-bit integers as \emph{Opcode}, all elements in VPA are 16-bit integers. 
This includes indexes to the VM. The VPA is a constant, global array. 
This allows any function to calculate the hash of any other virtualized function. 
Being declared as constant in LLVM also means we cannot generate code that modifies the array at runtime.

The decoded instruction's result will be written back into the VM to be used by subsequent instructions. 
The VM is allocated on the heap. 
When handling an instruction that returns execution to the caller, we clean up the allocated memory before returning (also see \Cref{subsubsec:data-array-restrictions}).

\subsubsection{Instruction Lifting and Translation}\label{subsubsec:translating-instructions}
In our design, we generate an interpreter for each virtualized function. 
Function level virtualization enables us to maintain the program's original function symbols and thus no linking problem will be incurred.
As our solution is implemented as a transformation pass based on LLVM, 
the maximum number of instructions we have to translate in our VO is limited by the maximum of instructions in LLVM, i.e. 58 distinct instructions.
\texttt{PHINode} is the only instruction that we cannot directly translate.
To cope with \texttt{PHINode} instructions, we resort to LLVM's \texttt{reg2mem} pass yielding replacement of \texttt{PHINodes} with corresponding allocations, load and store instructions.
Exceptions are not supported in the current implementation (see~\Cref{sec:unsupported-instructions}).  

Based on the user's list of sensitive functions, we sequentially virtualize each one individually. 
The virtualization step first iterates over all instructions in the current function. 
For each instruction visited, we generate a random 16-bit integer value to be used as the \emph{opcode}. 
If an instruction of this type has already been added, (granted that operands' types match) we reuse that instruction's \emph{opcode}. 
Next, for each operand, we add the value to VM and append the index to the VPA. 
For instructions that change the control flow, such as branch instructions, we append a placeholder instead.

After all instructions have been visited, we iterate over the translated instructions and, for instructions that change the control flow, 
replace the placeholders with correct indices of the branch destinations. 
Despite the flat nature of the VPA, we can still translate branch instructions. 
Given that LLVM branch instructions point to target basic blocks, we need to translate the target to an index in the VPA. 
The index should refer to the first instruction of the original destination block in the VPA. 
This is as simple as assigning the target instruction index to the VPC. 
Such assignments mimic jumps in programs.

For conditional branches, the situation is slightly different. 
Each conditional branch has two destinations depending on whether the condition holds true or false.
Therefore, we need to translate both target blocks to their corresponding indexes in the VPA and subsequently set the VPC accordingly. 
The branch translation step needs to be done after all instructions have been virtualized to guarantee that the index is correct. 
Otherwise, we might not find the correct instruction, if it has not yet been virtualized.

After all instructions have been virtualized, we add the current function with its hash to a temporary list.
This list is later used while crafting SC checkers (see \Cref{subsec:design-self-checksumming}). 
If the current function is a checker (dictated by the randomly created network of checkers~\Cref{subsubsec:network-of-checkers}), 
we insert a guard for each checkee at random locations into the VPA of the function.


\subsubsection{Interpreter}\label{subsubsec:interpreter}
An interpreter is generated for each individual function. 
Interpreter generation starts with allocating a VM and storing all needed constant values and function arguments into it. 
The VPC is initialized to zero. 
Next, a loop is added, which fetches the next instruction, decodes it and dispatches execution to the corresponding handler.
The interpreter starts with the instruction at index zero, 
and then increments the VPC by one for each fetched \emph{opcode} from the VPA.
In case of control-flow changing instructions, the VPC is set to indexes of target destinations in the VPA.

The main body of the interpreter is in fact a switch statement that takes the current \emph{opcode} as an argument. 
Each case in the switch body corresponds to a distinct opcode in the RISA.
Cases are \emph{handlers} of virtualized instructions. 
Handlers contain code that emulates the original instruction(s). 
This involves loading operands from the VM, executing the decoded instruction, and writing the result back into the VM.

Several instructions coded with the same opcode in the program will end up using the same handler.
In addition to reducing the binary size, reusing these handlers has several advantages. 
For one, there is no one-to-one mapping of handler and original instruction. 
This also implies that if an attacker wants to change how one specific instruction is handled, for example changing a jump destination, it will result in side-effects at different locations of the function. 
Therefore, it is intuitively harder to tamper with the program control-flow or, generally, instructions. 


Using function-level virtualization has the advantage that once a function's interpreter has been successfully reverse engineered, 
the attacker cannot transfer this knowledge to the other functions; instead, 
each function's interpreter has to be reverse engineered individually. 
Moreover, having virtualization at the level of functions yields a simpler linking process. 

Since the \texttt{ret} (return instruction in LLVM IR) can be emulated in the function-level virtualization,  
passing the return values between functions becomes trivial. 
However, a module-level interpreter would need to read and write the return values from a shared memory, 
which entails addressing challenges such as \emph{race conditions} and \emph{memory management} issues.


In our design, the default case of the interpreter's switch statement will invoke the response mechanism. 
That is we treat invalid opcodes as well as VPC issues as tampering attacks.

\subsubsection{Self-Checksumming Guards}\label{subsec:design-self-checksumming}
One way to tamper with the program behavior is to manipulate VPA values.
Our self-checksumming protection utilizes a set of code snippets, referred to as \emph{guards}, that verifies the VPA values of different (virtualized) functions.
In a nutshell, a guard hashes a function's VPA to ensure that it matches the expected value at different intervals during program execution. 

SC protection is added after virtualization. 
Guards obtain a unique opcode similar to other instructions. 
Each guard has a target function (checkee) as well as an expected hash value.
That is, for every guard opcode in VPA there exists two operands in VM.
Another memory slot is reserved in VM for the runtime hash value of guards.
During program execution, upon fetching a guard opcode the interpreter loads the address of the target function's (checkee's) VPA along with the expected hash from VM.
For the given target VPA a cumulative byte-by-byte hash is computed that is persisted in the preserved runtime hash in VM. 
The runtime hash is subsequently matched with the expected hash. In case of mismatches the response mechanism is triggered.
Bear in mind that depending on the desired protection configuration (see \Cref{subsubsec:connectivity,subsubsec:network-of-checkers}) 
more than one guard might be injected into the VPA of a given function.





As per the hash function, we use the binary \texttt{XOR} operator which comes with two clear benefits.
First, the operation occurs quite often in normal programs, e.g. to clear a register value. 
So, it offers a higher stealth.
Second, \texttt{XOR} is fast as it requires only a single operation on many X86 processors ~\cite{fog2011instruction-web}.
Nonetheless, our implementation can easily be extended to use different hash functions, if needed. 

\subsubsection{Guards Connectivity}\label{subsubsec:connectivity}
If a guard in function A checks the code of function B, then we call function A the \emph{checker} and function B the \emph{checkee}.
Connectivity refers to the number of checkers per each sensitive function. 
In the generated network, this is given by the number of incoming edges to a (sensitive) node. 
Since in our implementation we do not support cyclic checks, 
the connectivity is restricted by the number of functions in the module. 

\subsubsection{Network of Checkers}\label{subsubsec:network-of-checkers}

We generate a random network of checkers in the form of a directed acyclic graph. 
For each function in the sensitive set, we pick a number of functions equal to the desired connectivity to be checkers to the function of interest. 
In cases where the desired connectivity value is simply not achievable, for example because it is higher than the number of functions in the module, our pass will use the highest possible connectivity. 
This results in each sensitive function being checked by all other non-sensitive functions in the module. 


\subsubsection{Response Mechanism}
Hash mismatches in our solution are redirected to the default case in the interpreter's switch statement.
The response mechanism in our solution is straight-forward termination of the process by calling the C library function \texttt{abort}.
This can be extended to use a stealthier response (such as stack pollution) or a multitude of them, one of which is randomly selected at runtime.
Since the intrusiveness and hostility of the response mechanism depend on the use case (e.g. termination is not an options in browsers~\cite{banescu2015software}),
we let the users of the tool develop their desired responses.

\subsection{Optimization}\label{subsec:optimization}

The machine code that will be generated by our scheme is likely to be quite inefficient. 
The switch statement can, in the worst case (with no reuse of handlers), get as large as the number of instructions of the original function. 
This results in a multitude of comparisons and conditional branches. 
In some cases, the value that is used in the switch has to be compared against each case value until a match is found. 
Therefore, the generated code, in the worst case, has to iterate through plenty of cases and compare them to the given opcode value. 
Only on a match, will it jump to the specified block. Thereafter the same routine is repeated until no more instructions are left in the function's VPA. 

It is noteworthy to mention that some compilers reduce the number of comparisons by translating switch statements to indirect jumps through jump tables~\cite{bernstein1985producing}. 
This optimization can improve the performance at the cost of adding a multitude of jumps in the program binary.

Despite handler reuses and indirect jumps, the performance impact particularly for instructions within loops can be significantly high.
To cope with this limitation, we extend our interpreter to support indirect threading~\cite{interpreteroptimization}. 
This optimization enables directly connecting the handlers with each other, instead of having to iterate through the switch cases. 
The cornerstone of this technique is to compute the successor handler for each handler.
Afterward, a direct jump to the successor is placed at the end of handlers. 
If there exists more than one successor for a given handler, we compute the list of possible targets and insert an indirect branch. 
~\Cref{lst:obf-cf-opt} illustrates (using pseudo-code) how the control-flow looks like after applying this optimization. 
Note the \texttt{goto} instructions at the end of each handler.

\lstdefinestyle{interfaces}{
  float=t
}

\begin{lstlisting}[basicstyle=\footnotesize, style=interfaces, language=C,stepnumber=1,numbers=left,numberfirstline=false,caption={A simple demonstration of an interpreter utilizing the indirect threading optimization}, label={lst:obf-cf-opt}]
int some_func() {
  uint8_t *DataArray = ...; // store constants and function params
  uint64_t PC = 0; // program counter
  goto handler_1;
  handler_1: // this handler is being reused
    handle_op_1();
    PC += ...;
    next = CodeArray[PC]; // dynamically determine next handler
    goto handlers[next];
  handler_2:
    handle_op_2();
    PC += ...;
    goto handler_3;
    ...
}
\end{lstlisting}

In LLVM, indirect branches require a list of possible destinations and an address to jump to. 
While we can statically determine the list of destinations, we only know the correct addresses during execution. 
Not only are the actual addresses not calculated at the point of our pass, we are also using the handler for more than one instruction, 
so we also have more than one successor and need to account for multiple addresses. 
Since each instruction by itself only has one successor, we can simply use an index into the list of possible destinations and resolve the correct destination address at runtime. 
This is shown in listing~\ref{lst:obf-cf-opt} on lines 8 and 9.


Despite the potential performance improvements, we believe the optimization negatively impacts the security of the protection.
Connecting the handlers with each other also makes the resulting control flow more linear. 
Linear control flows makes the resulting machine code easier to analyze both for a human and static analysis tools. 
As a switch in machine code is nothing more than a sequence of comparisons and conditional jumps, symbolic execution tools quickly run into path explosion problems. 
Linear control flow for a human makes the sequence of executed instructions more intuitive than a series of comparisons and conditional jumps. 
Furthermore, the handlers without reuse can readily be spotted by attackers. 

Therefore, we keep this optimization optional, to be used in cases where large overheads are not tolerable.
In the remainder of this paper we refer to the unoptimized implementation as the \emph{secure} implementation. 


\section{Evaluation}\label{chap:evaluation}
In this section we conduct a set of experiments to measure the performance and to evaluate the security of our proposed scheme.
As pointed out earlier, our optimization is set to enhance the performance. 
However, the actual performance improvement as well as security implications of the \emph{optimized} approach are rather unknown.
Throughout this section we run our evaluations on both approaches to precisely capture advantages and disadvantages of each approach.




\subsection{Dataset}
To carry out our evaluations we use a subset of 22 programs from the MiBench suite~\cite{guthaus2001mibench} along with three open source CLI games, namely \texttt{tetris}\footnote{\url{https://github.com/troglobit/tetris}}, \texttt{snake}\footnote{\url{https://github.com/troglobit/snake}}, and \texttt{2048}\footnote{\url{https://github.com/cuadue/2048\_game}}, summing up to a total of 25 programs. 
It is worthwhile to mention that technical difficulties in compiling some MiBench programs to a single LLVM IR bitcode file 
(e.g. use of a mix of assembly and c code) forced us to exclude them from our evaluations.
The first four columns in \Cref{tbl:resilience-cov100} show details (\#Instruction: number of instructions, \#Function: number of functions, and \#Block: number of basic blocks in LLVM IR) for each program in our dataset. 

\begin{table}[t]
	\centering
	\begin{tabular}{|l|l|l|l|l|l|l|}
		\hline
		\rot{\textbf{Program}} & \rot{\textbf{\#Inst.}} & \rot{\textbf{\#Function}} & \rot{\textbf{\#Block}} & \rot{\textbf{\#Prot. Inst.}} & \rot{\textbf{Prot. Inst. \%}} \\ \hline
		qsort\_s               & 107                    & 2                         & 18                     & 74                           & 69                            \\ \hline
		crc                    & 152                    & 4                         & 20                     & 135                          & 89                            \\ \hline
		qsort\_l               & 166                    & 2                         & 22                     & 128                          & 77                            \\ \hline
		dijkstra\_l            & 338                    & 6                         & 59                     & 304                          & 90                            \\ \hline
		dijkstra\_s            & 338                    & 6                         & 59                     & 304                          & 90                            \\ \hline
		rawcaudio              & 437                    & 3                         & 78                     & 389                          & 89                            \\ \hline
		rawdaudio              & 437                    & 3                         & 78                     & 389                          & 89                            \\ \hline
		basicmath\_s           & 538                    & 5                         & 63                     & 186                          & 35                            \\ \hline
		basicmath\_l           & 649                    & 5                         & 83                     & 186                          & 29                            \\ \hline
		sha                    & 666                    & 8                         & 57                     & 619                          & 93                            \\ \hline
		tetris                 & 669                    & 13                        & 129                    & 564                          & 84                            \\ \hline
		bitcnts                & 705                    & 15                        & 82                     & 669                          & 95                            \\ \hline
		fft                    & 760                    & 7                         & 91                     & 483                          & 64                            \\ \hline
		2048                   & 803                    & 17                        & 146                    & 699                          & 87                            \\ \hline
		search\_l              & 873                    & 10                        & 159                    & 741                          & 85                            \\ \hline
		search\_s              & 873                    & 10                        & 159                    & 741                          & 85                            \\ \hline
		snake                  & 1124                   & 13                        & 172                    & 1071                         & 95                            \\ \hline
		patricia               & 1201                   & 6                         & 169                    & 867                          & 72                            \\ \hline
		bf                     & 3667                   & 8                         & 168                    & 3383                         & 92                            \\ \hline
		rijndael               & 5924                   & 7                         & 147                    & 5074                         & 86                            \\ \hline
		say                    & 7447                   & 75                        & 1302                   & 7135                         & 96                            \\ \hline
		susan                  & 12996                  & 19                        & 916                    & 12760                        & 98                            \\ \hline
		toast                  & 15374                  & 94                        & 1542                   & 15268                        & 99                            \\ \hline
		djpeg                  & 54496                  & 379                       & 6518                   & 53204                        & 98                            \\ \hline
		cjpeg                  & 56735                  & 391                       & 6788                   & 55396                        & 98                            \\ \hline\hline
		\textbf{Mean}          & 6699.00                & 44.32                     & 761.00                 & 6430.76                      & 83                            \\ \hline
		\textbf{Median}        & 760.00                 & 8.00                      & 129.00                 & 669.00                       & \textbf{89}                            \\ \hline
		\textbf{Std}           & 15253.74               & 104.82                    & 1817.06                & 14937.38                     & 18                            \\ \hline
	\end{tabular}
	\caption{Protection coverage of instructions including the network of checkers using SC atop VO (\emph{Prot. Inst.\%} column)}\label{tbl:resilience-cov100}
\end{table}

\subsection{Coverage}
The goal of this experiment is to determine the effectiveness of the protection scheme.
One way is to measure the protection coverage.
Coverage of protection in this case refers to the number of instructions that are virtualized as well as the number of instructions (in the VPA) that are protected by SC guards.

We generated a set of protected binaries (VO+SC) with a function coverage of 100\%. 
A coverage of 100\% indicates that the protection deems all the program instructions as security sensitive and thus tries to protect them all.
Since the checkers of SC guards are randomly selected, we repeat the binary generation 20 times to weed out potential noises yielding a total of 480 programs, 20 $\times$ 24 (programs).
Bear in mind that utilizing optimizations has no impact on the coverage of SC.
Columns \#Prot. Inst. and Prot. Inst.\% of ~\Cref{tbl:resilience-cov100} capture the number and percentage of SC protected instructions, respectively. 
\Cref{sec:protection-coverage} discusses the impact of the protection coverage on the scheme security.



\subsection{Performance}\label{sec:performance-evaluation}
The goal of this experiment is to measure the overhead of the added protection in both secure and optimized modes. 
We intend to capture the actual overhead of VO in both modes. 
Then, we repeat the same set of experiments for the combination between VO and SC.
This way of measurement enables us to single out the overhead of each scheme separately.

To measure the impact of our protection, we generate protected binaries with a range of partial protections (i.e. 10\%, 20\%, and 50\% of functions) as well as a complete protection (i.e. 100\% of functions).
For every protection level we generate 20 random combinations of functions to be deemed as sensitive. 
Thereafter,  if the SC protection is enabled, we generate 10 protected programs for each combination to weed out the noise of SC's random network of checker creation.
That is, for the VO+SC benchmarks we end up with 800 (20 [function combinations] $\times$ 4 [protection levels] $\times$ 10 [network of checkers]) protected instances for each mode (secure and optimized).
However, when only VO is applied we generate 80 ($20 \times 4$) protected instances per mode.
It is noteworthy that we set \texttt{connectivity=2} for SC throughout our experiments.
That is, every sensitive functions is checked by 2 other functions, if enough functions are available in the module.

We use \texttt{benchexec}\footnote{\url{https://github.com/sosy-lab/benchexec}}~\cite{beyer2016reliable} to precisely measure the performance of our programs before and after applying protections. 
We run each program 100 times with the exact same inputs and measure the overhead. 
For games we pipe in constant input by intercepting the necessary library calls (e.g. \texttt{getch}).
According to our experiments all the protected programs execute correctly with respect to the provided inputs.


\subsubsection{Secure Mode}
\begin{figure*}[t!]
	\centering
	\includegraphics[clip, trim=0.25cm .375cm 0.3cm 0.2cm,width=0.95\textwidth]{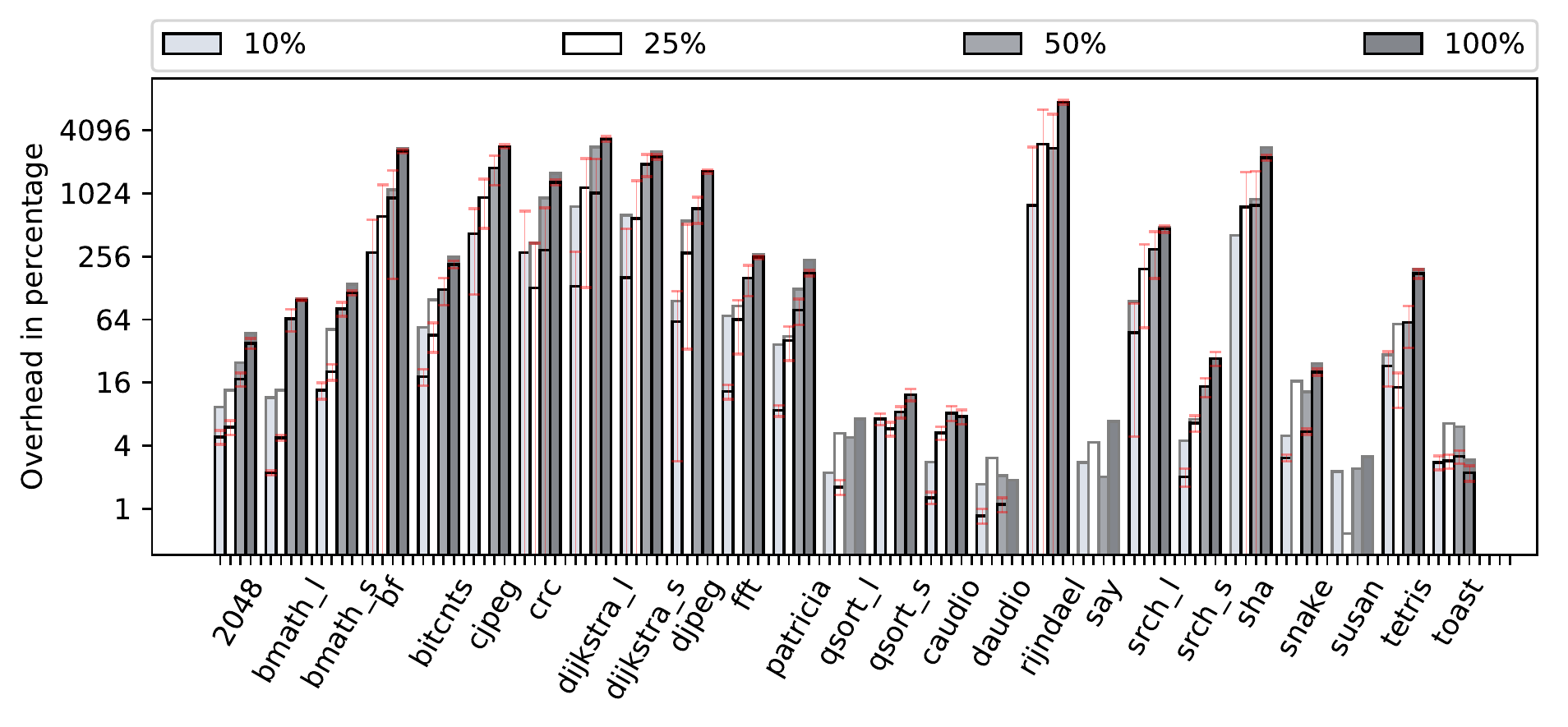}
	\caption{Overhead of VO vs. VO+SC protection in the secure mode; the gray-margined bars capture VO+SC protection while the black-margined bars depict VO protection results}
	\label{fig:overhead-VO}
\end{figure*}

\Cref{fig:overhead-VO} illustrates the overhead of VO as well as VO+SC in the secure mode . 
The average overhead of VO (applied single handed) is 89.34\%, 314.95\%,  445.41\%, and 1018.46\% for protection levels of 10, 25, 50 and 100, respectively.
When VO and SC are combined the average overhead rounds to 109.74\%, 193.64\%, 468.72\%, and 1055.51\% for protection levels of 10, 25, 50 and 100, respectively.
Note that the average of VO+SC overheads for the 25\% protection level are smaller than the corresponding overheads when only VO is applied.
We believe this is due to the randomness of function selections, the frequency of check executions, and potential LLVM optimizations.

\begin{figure*}[t!]
		\centering
		\includegraphics[clip, trim=0.25cm .375cm 0.3cm 0.2cm,width=0.95\textwidth]{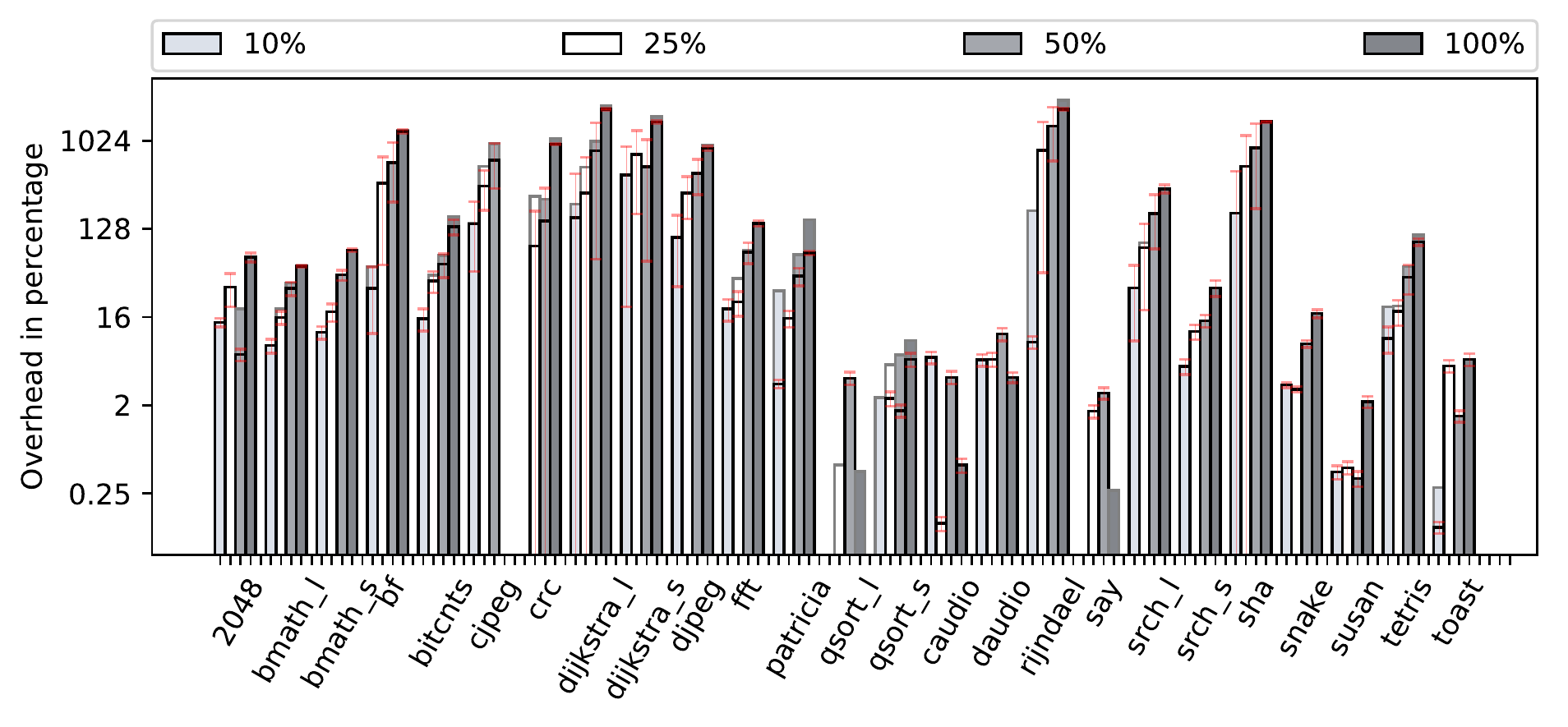}
	\caption{Overhead of VO vs. VO+SC in the optimized mode; the gray-margined bars capture VO+SC protection while the black-margined bars depict VO protection results}
	\label{fig:overhead-VO+SC}
\end{figure*}

%
%

\subsubsection{Optimized Mode}
In this section we report on the average overhead of our protection in the optimized mode. 
\Cref{fig:overhead-VO+SC} captures the results of our experiments. 
In sum, the induced overhead for VO-only is 48.27\%, 151.59\%, 231.65\%, and 458.42\% for protection levels of 10, 25, 50 and 100, respectively.
For VO+SC protection, the captured results indicate an average overhead of 32.64\%, 150.54\%, 231.65\%, and 501.04\% for protection levels of 10, 25, 50 and 100, respectively.

It it noteworthy to mention that some protected (VO+SC) instances (in both secure and optimized modes) perform better than the unprotected version (VO only). 
We investigated those binaries manually but were not able to find any problems in them.
They appear to benefit much more from the LLVM optimizations compared to other protected binaries.

\subsection{Security}\label{sec:security-evaluation}
We analyze threats to the security of our protection from three perspectives: threats to SC protection, threats to VO, and threats to the combined protection. 
The utilization of \emph{secure} or \emph{optimized} approaches naturally impacts the protection level.
Since our optimization is merely applied on the interpreter of VO, the other components are not impacted.  
Therefore, we make a distinction between the two approaches only in the security analysis of VO. 
\subsubsection{Threats to SC protection}\label{subsec:attacks-on-sc}
\begin{figure}[t]
	\centering
	\includegraphics[width=0.5\textwidth]{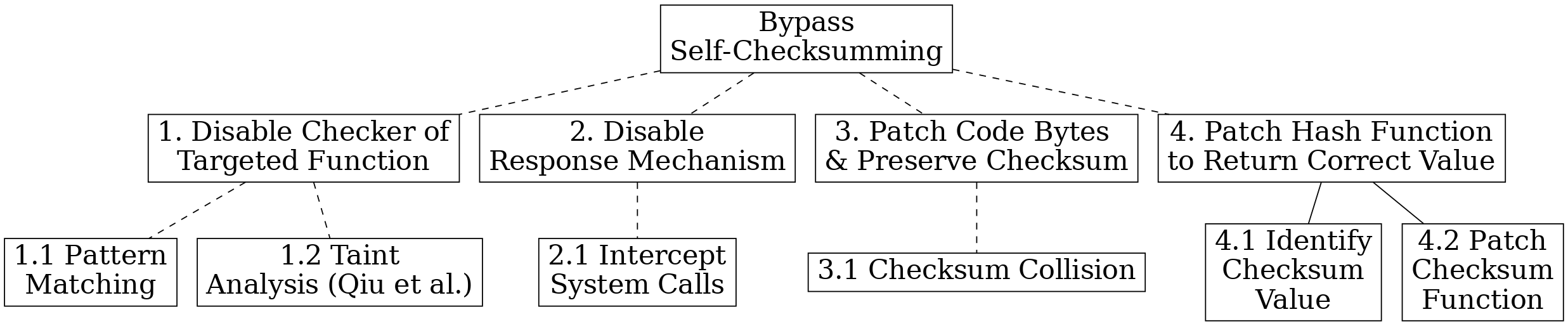}
	\caption{Attacks on SC depicted in the form of attack tree notation~\cite{banescu2017detecting} }\label{fig:attack-tree-sc}
\end{figure}

Disabling the self-checksumming is one way to tamper with a protected program. 
~\Cref{fig:attack-tree-sc} represents attacks on SC protection~\cite{banescu2017detecting} in the form of an attack tree.  
Dashed lines and solid lines stand for disjunction and conjunction refinements.
The following paragraphs describe each of the nodes at depth one in more detail.

\paragraph{Disable Checkers}
An attacker may disable the checkers for the targeted function. 
The first difficulty in this case is identifying all guards in functions that are checkers. 
If attackers successfully identify such checkers, they still might need to identify checkers covering the target checkers (overlapping guards).
Since we utilize VO at the function level, 
attackers have to analyze and reverse engineer the interpreters of all checkers individually. 
Therefore, the complexity and the amount of work needed for a successful attack is considerably higher.

~\citeauthor{qiu2015identifying} ~\cite{qiu2015identifying} present a generic approach for defeating SC protection using taint analysis.
However, as also pointed out by the authors, such attacks fail to affect our protection due to the utilization of VO. 
Their attack identifies the verification of SC hashes (conditional branch) in a given program. 
Thanks to VO multiple branches (not only SC verifications) may reuse the same handler rendering the attack ineffective.

It is worthwhile to mention that we consider attacks that require a modification in the OS (such as~\cite{wurster2005generic}) 
limited for mass distributions and therefore less critical.
\paragraph{Disable Response Mechanism}
Disabling the response function is unlikely to be successful, but possible. 
After detecting a hash mismatch, the interpreter/guard branches to a specific handler and calls the \texttt{abort} function. 
If an attacker were to disable this function call, 
the program will likely run into a segmentation fault or, at the very least, undefined behavior. 
Mainly because the value of \emph{VPC} in such cases is not properly set. 

Completely removing the call to the response function from the handler 
or changing the jump address after the comparison of hash values
are considered as an attack against the interpreter. 
These attacks are explained in section~\ref{subsec:attacks-interpreter}.

\paragraph{Patch Code Bytes \& Preserve Checksum}
VO tends to significantly hinder this attack. 
For machine code it is possible to find a sequence of instructions that perform a useful task, 
while hashing to the same expected hash.
However, the fact that the changes need to go through the interpreter further limits the range of possibilities.

\paragraph{Patch Hash Function}
An attacker can modify the hash function to always return the correct value for the current checkee. 
The actual hash value is stored in the function VM.
Therefore, reverse engineering VO, particularly the interpreter's accesses to VM, is a prerequisite for the success of such attacks. 
To find the correct location where the precomputed checksum is stored, it is also necessary to reverse engineer the interpreter, which we discuss in the following section.


\subsubsection{Threats to VO}\label{subsec:attacks-on-virt}
It is important to make a distinction between the approaches (namely \emph{optimized} and \emph{secure}) when discussing threats to VO.
The optimized approach (without further hardening) stands worse chances against attacks. 
Mainly because VPC is replaced with direct jumps to the next block(s). 
This on the one hand enhances the performance.
On the other hand, the optimization eases the control flow recovery of the obfuscated code using static analysis.
Utilizing resilient opaque predicates can enhance this situation.
We believe the rest of security analysis remains indifferent for both approaches.
Therefore, we refer to both approaches as VO in the remainder of this section. 

Attacks on VO can be classified into two categories: \textbf{i)} generic attacks on VO; and \textbf{ii)} manual attacks on the interpreter. 
\paragraph{Generic attacks on VO} Table~\ref{tbl:attack-vectors} summarizes a few state-of-the-art attacks on VO along with their advantages and disadvantages.

The approach by \citeauthor{kinder2012towards}~\cite{kinder2012towards} requires knowledge about the VPC used in the interpreter. 
They assume there exists a function to reliably detect the storage location of the VPC. 
After our transformation (in the program binary representation) we cannot reliably make assumptions about the location of the VPC. 
Compiling with different levels of optimizations (e.g. \texttt{-O3}) might reserve a register value for it, but it might also end up being written to and loaded from memory.

\citeauthor{coogan2011deobfuscation}~\cite{coogan2011deobfuscation} propose to use taint analysis in order to identify the relevant instructions, i.e.~instructions that affect the values of program outputs in a system trace.
This approach is able to identify and remove interpreter instructions from traces.
However, it does not remove the SC guards.
Moreover, the output of this approach is a simplified trace for each different input, which does not tell the attacker how to identify and disable the SC guards from the protected program.

\citeauthor{yadegari2015generic}~\cite{yadegari2015generic} significantly improve the attack by \citeauthor{coogan2011deobfuscation}, by more advanced trace simplification techniques, symbolic execution and recombining the simplified traces into a simplified control flow graph (CFG). 
The drawback of this approach is the large input space that the protected program may accept as input, because in order to reconstruct an accurate CFG, all possible paths in the code must be exercised first.
Even if this is achieved, the attacker still has to detect and remove the SC guards, which requires significant effort overall.

\citeauthor{salwan2018symbolic}~\cite{salwan2018symbolic} developed a novel approach to defeat VO by combining taint analysis, symbolic execution and code simplification. 
Compared with known attacks that target the control-flow graph, the output of their script~\footnote{\url{https://github.com/JonathanSalwan/Tigress_protection}} is expected to be a clean devirtualized executable. 
Their technique also symbolizes the VPC.
We were unable to successfully use their tool against executables protected with our protection, due to an error during execution. 
We have already reported the bug but have not yet heard back from the authors.
Nevertheless, we believe that attacking our approach will run into the path explosion problem for the real-world programs in our dataset. 


\begin{table}[t]
	\begin{threeparttable}
		\centering
		\resizebox{\linewidth}{!}{
		\begin{tabular}{|c|l|l|l|l|}
			\hline
			& Kinder~\cite{kinder2012towards} & Coogan~\cite{coogan2011deobfuscation} & Yadegari~\cite{yadegari2015generic} & Salwan~\cite{salwan2018symbolic} 																																\\ \hline
			\begin{tabular}{c}\textbf{Attack}\\\textbf{Type}\end{tabular} 
			& \begin{tabular}{l}SA\tnote{1}\end{tabular} 
			& \begin{tabular}{l}TA\tnote{2}\end{tabular} 
			& \begin{tabular}{l}bit-level TA\tnote{2}\end{tabular} 
			& \begin{tabular}{l}SA\tnote{1}\ \ \& TA\tnote{2},\\formula simpl.,\\code simpl.\end{tabular} 
			\\ \hline
			\begin{tabular}{c}\textbf{Attacker}\\\textbf{Goal}\end{tabular}  
			& \begin{tabular}{l}approx.\\data values\end{tabular}				  				 
			& \begin{tabular}{l}significant\\trace\end{tabular} 											
			& \begin{tabular}{l}\emph{CFG}\end{tabular}				 
			& \begin{tabular}{l}extract\\original code\end{tabular}																																									
			\\ \hline
			\textbf{Output} 			 																					 
			& \begin{tabular}{l}\emph{CFG}\\and invariants\end{tabular} 				
			& \begin{tabular}{l}simpl. trace\end{tabular}  		 											
			& \begin{tabular}{l}simpl. \emph{CFG}\end{tabular}	 
			& \begin{tabular}{l}simpl. code\end{tabular} 																																													
			\\ \hline
			\textbf{Drawback} 																							 
			& \begin{tabular}{l}assumptions\\on interpreter\\structure\end{tabular} 
			& \begin{tabular}{l}equation\\to \emph{CFG}\\conversion\end{tabular} 
			& \begin{tabular}{l}large\\input space\end{tabular}   
			& \begin{tabular}{l}path explosion,\\symb. reasoning\end{tabular} 
			\\ \hline
		\end{tabular}
	}
		\begin{tablenotes}
			\item[1] Static Analysis
			\item[2] Taint Analysis
		\end{tablenotes}
	\end{threeparttable}
	\caption{Known attacks on virtualization obfuscation~\cite{banescu2016vot4cs, salwan2018symbolic}}\label{tbl:attack-vectors}
\end{table}

\paragraph{Attacks on the Interpreter}\label{subsec:attacks-interpreter}
In addition to known attacks against VO, 
it is also possible to attack the interpreter itself without reverse engineering the executed bytecode. 
In the executable, it is possible to change the jump address of a specific block to another location. 
This can happen in functions with few branches, where the handler for the branch instruction is not reused. 
We try to mitigate this by reusing handlers and thus increasing side effects when a handler has been tampered with. 
Changing the address of a jump will then be changed for \textit{all} branches using this specific handler. 
This effect can further be increased by adding opaque predicates, 
generating even more branches and increasing the number of instructions using the same handler.

Identifying a handler that is only used by a single instruction in a function not being checked seems to be one of the most promising attack scenarios.
However, this means that the instruction is not a sensitive instruction and may not be useful for the attacker to tamper with.

In general, the interpreter can be hardened by other protection techniques like any other piece of software. 
This is also one of the main reasons we have decided to implement our solution using LLVM: 
to increase the composability with other, existing software protection techniques.

\subsubsection{Threats to the Combined Protection}
As mentioned in section~\ref{subsec:attacks-on-sc}, attacking SC requires reverse engineering the virtualization obfuscation. 
SC is implemented on top of VO and thus protected by it.
Changing \emph{opcodes} in the generated bytecode can be identified by having SC guards in place. 
We can conclude that it is necessary to first reverse engineer our VO, in order to defeat SC guards. 
Ignoring these protections and attacking the interpreter itself, as outlined earlier generates side-effects across each function due to the reuse of handlers.


\section{Discussion}

This section discusses tradeoffs between the performance and the security of our approach. We conclude this section by discussing how the security of our approach could be improved.

\label{sec:discussion}
 \subsection{Protection Coverage}\label{sec:protection-coverage}
 Our results confirm that our implementation is capable of virtualizing all instructions in the given programs. 
 The SC protection on top of VO is capable of protecting a large portion of instructions (a median of 89\%) including the network of checkers.
 Some instructions however were left unprotected. 
 Our investigations indicate that such instructions reside either in the root of the network of checkers or in the nodes with no incoming edges. 
 Since we do not support cycles, some nodes might be left unprotected. 
 Bear in mind that we can easily achieve 100\% coverage over the original program instructions by not checking the guards themselves to avoid cycles.
 We believe not checking guards downgrades the resilience of the protection. 
 Nonetheless, supporting cycles is a matter of engineering efforts.

\subsection{Performance vs. Security}
As with every software protection, there is a trade off between performance and security. 
Adding security, for example in the case of our secure interpreter, results in additional code to be executed. 
The extent to which added protections impact the performance depends on various factors such as the number of comparisons, loops, etc.


As showed in the performance evaluation~\ref{sec:performance-evaluation}, our secure implementation yields on average an overhead of 1055.51\%. 
Being $\approx$ 10x slower than the unprotected executable is on the upper end of worst case performances for regular interpreters. 
However, this approach is also hard for symbolic execution engines to analyze, due to path explosion. 
An important remark here is the low overhead that SC protection, with a connectivity of 2 checkers per checkee, imposes on the protected binaries. 
For the 25\% protection level the VO+SC protected binaries outperformed VO-protected binaries. 
As pointed out earlier, the randomness of the network of checkers appears to be the cause of this phenomenon. 
It is worth mentioning, that the interpreter in this solution has no optimizations whatsoever applied to it.

Further research needs to be conducted to develop and subsequently utilize a set of potential optimizations that do not downgrade security. 

The optimized implementation, which utilized indirect threading in the interpreter, resulted in far more efficient protected binaries.
Our results indicate an approximate decrease of $\approx$ 50\% in the overhead of protection.
More importantly, in the optimized mode, the contribution of SC protection to the overhead is significantly decreased. 
That is, in 10, 25, and 50\% protection levels SC did not impose any extra overhead but overheads dropped.
However, one can not conclude that utilizing SC yields better performance as other factors such as randomness of checkers and LLVM optimizations need to be taken into account. 
For the 100\% protection level SC constitutes only $\approx$ 43\% of the protection overhead (501.04\% VO+SC - 458.42\% VO only). 
It appears that LLVM manages to significantly optimize the SC protection.
A large portion of the reported overheads stem from  LLVM's inability to optimize VO. 
The use of VM introduces a complexity in the framework's analysis as it transforms direct variable accesses into dynamic array lookups. 
This makes it difficult for LLVM to analyze control and data flow and thus hinders further native optimizations in IR. 
Besides, the VM itself imposes indirect accesses by storing and subsequently retrieving values from memory instead of registers, which substantially slows down the execution.
In the case of the secure version, the switch statement is translated to multiple comparison and jump instructions, which decreases the effectiveness of the branch prediction. 
As the value to switch over is loaded dynamically, this can not be improved by static analysis nor compiler optimizations. 
Instruction prefetching in these cases is also impacted due to the number of jump instructions generated by the switch statement.

We believe as the optimization removes the switch statement and connects each handler with its successors,
it becomes easier for LLVM to analyze and run additional optimizations.
Despite the performance gain, the indirect threading optimization makes the generated code easier for a human to analyze. 
Removing the switch also reduces the amount of possible paths allowing more efficient symbolic execution based attacks.



Lastly, the introduced performance overheads heavily depend on the nature of the program to protect. 
The structure of the program's call graph, number of nested loops, and the degree of IO dependency play important roles in the overall overhead of the protection.

\subsection{Security}

By using a switch statement to dispatch all opcodes, we were able to force symbolic execution engines to run into the problem of path explosions. Even if the original program did not contain many different paths, after transformation every case in the switch statement corresponds to one path. The highest possible number of cases in the generated switch statement is equal to the number of instructions in the original function. 
This happens when we cannot reuse a single handler. 
Usually, certain instructions appear more than once, for example to load a value. 
We try to reuse as many handlers as we can to increase side-effects of attacking the interpreter (see~\Cref{subsec:attacks-interpreter}). 
This heavily depends on the structure of the original module, for example the number of instructions in functions. 
The amount of possible paths can be further increased by introducing opaque predicates~\cite{fedler2014isa} in our interpreter.

Our protection can fall short when an attacker can identify a handler that is neither reused, nor protected by SC. 
Supporting cyclic checks in SC can mitigate this kind of attack. 
One great benefit of our approach is that  any type of LLVM-based software protection can be employed on top or before our transformation to increase the resilience of the protection.

\section{Limitations}
\label{sec:limitations}

This section presents limitations of our approach and implementation. Several of these limitations can be overcome and we plan to address them in future work.

\subsection{Unsupported Instructions}\label{sec:unsupported-instructions}
There are a few Windows-specific instructions regarding exception-handling that we do not support. 
This is in part due to the fact that Windows does not allow for loadable modules, and our infrastructure is almost exclusively based on *nix. 
Since our pass is loaded via LLVM's \texttt{opt} tool, we would need to integrate the pass into LLVM's internal optimization pipeline. 
Doing so increases development time significantly, because every change made to the pass results in having to rebuild a lot of LLVM's tools. 
We have decided to move support for Windows to future work.

\subsection{VM Restrictions}\label{subsubsec:data-array-restrictions}
Calling \texttt{free} on VM can be problematic, when exceptions are involved. 
This is a well-known source of memory leaks. 
If a called function were to throw an exception that is not caught, the VM will never be cleaned up. 
In C++, this is typically solved with a technique called \emph{RAII}~\cite{cpprefraii}, but we do not have access to this in LLVM IR. 
Therefore, in these special cases, our implementation is leaking the memory of the VM. This is an issue that we hope to resolve in future work.

There are instructions in LLVM that require their operands to be constant, meaning they have to be known at compile time. 
Consequently, we are unable to load those dynamically from the VM. 
An example for this is the intrinsic LLVM \texttt{memcpy} function. Two of its parameters, \textit{alignment} and \textit{volatile}, are required to be constant. 
Due to the fact that values loaded from the VM at runtime are not constant, in these cases we reuse the original instruction's operands.

Another restriction of using a VM is that we cannot add global variables to it. Since all of our generated instructions operate solely on the VM, we would not assign to the global variable. 
Calling another function that depends on the updated value would change the program's behavior. 
This could, in theory, be solved by updating both the global variable, as well as the VM. 
However, calling a function that updates the global variable would require us to update the value in the VM. 
This requires performing non-trivial control flow analysis to determine read and write accesses to the global variable.
A simple solution to this issue is to not add global variables to the VM, but instead simply use the global variable itself. 

\section{Conclusions}
\label{sec:conclusions}
We designed, implemented and evaluated an LLVM-based protection that effectively combines virtualization obfuscation (VO) and self-checksumming (SC). 
Our approach removes the need for binary post-patching of SC pre-computed hash (checksum) values. 
As a direct consequence, SC is completely architecture-agnostic and better yet, fully composable with other protection techniques.
More importantly, our SC scheme benefits from the hardening added by VO and thus exhibits a higher resilience.

Regarding VO, we presented and later evaluated two implementations, namely \emph{secure} and \emph{optimized}.
We conducted a set of performance as well as security evaluations, where a set of 25 real-world programs (MiBench and CLI games) were used.
The secure version stands better chances against symbolic execution attacks.
However, it imposes an average overhead of $\approx$10x for full protection (i.e.~100\% of instructions are protected), of binaries.
In contrast, the optimized version imposes an average overhead of 5x for the same protection level.
The SC protection itself imposes only 43\% overhead for full protection with a connectivity of two.
We believe such overheads are acceptable when protection is applied on a subset of sensitive segments of programs (e.g. license checking).


In the security evaluation we discussed attacks on SC and VO. 
Possible mitigations were also explained in our evaluations. 
Since our solution adds SC on top of VO, 
an attacker has to first break the obfuscation, in order to be able to attack SC.
To the best of our knowledge, there exists no tool that can automatically carry out such attacks. 
Therefore, a combination of tools need to be manually applied by attackers to break the scheme. 

As future work we plan on implementing cyclic checks for SC. 
Another interesting area is to further investigate means to optimize VO without (or with less) security sacrifices. 






%
%
%
 \bibliographystyle{ACM-Reference-Format}
 \bibliography{main}

\end{document}